\documentclass[a4paper]{jpconf}
\usepackage{graphicx}
\begin{document}
\title{Spacetime Brout--Englert--Higgs effect in General Relativity interacting with  p-brane matter\footnote{Contribution to the Procs. of the ERE2009 Meeting, Bilbao, September 2009}}

\author{Igor A. Bandos}

\address{Ikerbasque, the Basque Foundation for Sciences, and Department of Theoretical Physics and History of Science, the Basque Country University, P.O. Box 644, 48080 Bilbao, Spain\footnote{Also at ITP NSC KIPT, Kharkov, Ukraine}}

\ead{igor\_bandos@ehu.es}

\begin{abstract}
We review the manifestation of the Brout-Englert-Higgs effect in general relativity interacting with point-like and extended objects ($p$--branes including  string for $p$=1 and membrane for $p=2$), which manifests itself in the appearance of the brane source in the Einstein equation while the  graviton remains massless \cite{BdAI=PRD02}--\cite{IB+JdA=JHEP05}, and discuss briefly its relation and differences with the model for massive spin 2 field proposed recently by G. t'Hooft \cite{t'Hooft=07}.
\end{abstract}

\section{Introduction}

Brout-Englert-Higgs effect \cite{Higgs:1964ia}, which was also known as Higgs effect, consists in that the gauge fields of an internal symmetry group acquires the mass when the gauge symmetry is spontaneously broken. The increasing of the number of polarizations is explained by that the gauge fields ''eat'' Goldstone fields of the spontaneously broken gauge symmetry and incorporates their degrees of freedom as additional polarizations.

To be more specific, let us consider, following \cite{t'Hooft=07}, the following Lagrangian
\begin{eqnarray}\label{LHiggs}
& {\cal L}= - tr \left( {1\over 4} F_{\mu\nu}F^{\mu\nu} +{1\over 2} D_{\mu}\varphi^s
D^{\mu} \varphi^s + V(\varphi^2)\right)\;  \qquad
\end{eqnarray}
for the gauge field ${\cal A}_\mu={\cal A}_\mu^s T_s$ and the scalars $\phi=\psi^sT_s$ in the adjoint representation of the gauge group.
If the potential $V(\phi)$ forces the scalar field to have a non-vanishing vacuum expectation value $<\phi >$, one can fix the `preference gauge' \cite{t'Hooft=07} $\phi = <\phi >$. In this gauge the kinetic term for the scalar field- Goldstone fields for the spontaneously broken gauge symmetry- gives rise to the mass term of the gauge field,
${1\over 2} D_{\mu}\varphi^sD^{\mu}\varphi^s \longmapsto \propto m^2 {\cal A}_\mu {\cal A}^\mu $ with
$m^2 \propto <\phi^2>$, and
the scalar field equations are reduced to $\partial^\nu {\cal A}_\nu=0$. These equations are dependent: they  can be also obtained
as a selfconsistency conditions for the gauge field equations
 \begin{eqnarray}\label{DF=mA} D^\mu F_{\mu\nu} =  m {\cal A}_\nu\; , \qquad m^2 \propto <\varphi^2>\;   . \qquad
\end{eqnarray} This fact reflects the pure gauge nature of the scalar field in (\ref{LHiggs}).

{\it General Relativity} is invariant under the spacetime {\it diffeomorphisms}
(general coordinate transformations) $Diff_4$ and the graviton  can be identified with the gauge field for the $Diff_4$ gauge symmetry. Then it was natural to expect that the spontaneous breaking of diffeomorphism symmetry should result in that the graviton becomes massive, see \cite{Duff=PRD75,t'Hooft=07} and refs. in
\cite{t'Hooft=07,BdAI=PRD02}.

On the other hand, the spontaneous breaking of diffeomorphism symmetry occurs in the presence of material objects: particles (0-branes), strings (1-branes), membranes (2-branes) and p-branes with $p>2$ for higher dimensional cases $D\geq 4$. The corresponding spacetime Higgs effect, its supersymmetric generalization and consequences were the subjects of study in  \cite{BdAI=PRD02,BdAIL=PRD03,BdAIL=Higgs03,IB03-ProcsJdA} and in \cite{IB+JdA=JHEP05} (where some statements of previous papers on brane degrees of freedom were refined/improved). This spacetime Higgs effect manifests itself in the modification of the `free' Einstein equations by the $p$-brane sources, {\it i.e.} by the energy-momentum tensor localized on the $p$-brane worldvolume $W^{p+1}$,
\begin{eqnarray}\label{EiEq=gauge}
& {\cal G}^{\mu\nu}(g):= \sqrt{|g|}
({\cal R}_{\mu\nu}- {1\over 2} g_{\mu\nu}{\cal R}) = {T_p\over 4} \sqrt{|g^{[p+1]}|}\,g^{[p+1]mn}(\xi, \vec{0})
\delta_m^\mu\delta_n^\nu \delta^{(D-p-1)}(x^I)\,   \qquad
\end{eqnarray}
(see below for the notation). However, although such an energy-momentum tensor is a counterpart of the mass term in (\ref{DF=mA}), it cannot be considered as a mass term for graviton but rather as a counterpart of the cosmological constant contribution to the graviton equation \cite{BdAI=PRD02}-\cite{IB03-ProcsJdA}.

The above statements of \cite{BdAI=PRD02}--\cite{IB03-ProcsJdA} might seem to be in contradiction with \cite{Duff=PRD75} and particularly with \cite{t'Hooft=07}, where a model looking similar to brane interacting with gravity is discussed. The aims of the present contribution is to review the pure bosonic results of the study in   \cite{BdAI=PRD02}--\cite{IB03-ProcsJdA} and to point out the differences between gravity interacting with branes and the model considered in \cite{t'Hooft=07} which result in different properties of spin $2$ fields in these dynamical systems.

\section{Spacetime Higgs effect in the presence of p-brane matter}

In the case of ($D$-dimensional) general Relativity, the gauge symmetry is diffeomorphism symmetry $SDiff_D$ 
and the gauge field can be identified with metric
$g_{\mu\nu}= e_\mu{}^a(x)\eta_{ab}e_\nu{}^b(x)$,
\begin{eqnarray}\label{dgcg=}
\delta x^\mu := x^{\mu\prime}- x^\mu=- a^\mu(x)\; , \qquad  \delta g_{\mu\nu}:=  g_{\mu\nu}^\prime (x)-  g_{\mu\nu}(x)=
 \partial_\mu a^\rho  \, g_{\rho \nu}+
\partial_\nu a^\rho  \, g_{\mu\rho} +
 a^\rho  \partial_\rho g_{\mu\nu} \; .
\end{eqnarray}
Material objects, $p$-branes (particles for $p$=$0$, strings for $p$=$1$)
can be described by the coordinate functions $\hat{x}^\mu(\xi^m)$  defining parametrically their worldvolume $W^{p+1}$ as a surface in spacetime $M^D$,
\begin{eqnarray}\label{WinM}
{\cal W}^{(p+1)}\subset M^{D}\quad : \qquad x^\mu =\hat{x}^\mu (\xi^m)=\hat{x}^\mu (\tau , \vec{\sigma}) \; , \qquad \{\; {}^{\mu=0,1,\ldots, (D-1)\; ,} _{\quad m=0,1,\ldots, p\; .} \quad
\end{eqnarray}
Their transformations under $SDiff_D$
 \begin{eqnarray}\label{vdx=}
\delta \hat{x}^\mu (\xi) := \hat{x}^{\mu\prime}(\xi) - \hat{x}^\mu(\xi) =a^\mu(\hat{x}(\xi)) - \delta \xi^m \partial_m \hat{x}^\mu (\xi) \;  \qquad
\end{eqnarray}
indicate their Goldstone nature. This is not surprising as far as in flat spacetime (superspace) the $p$-branes break spontaneously the rigid translation symmetry of the bulk spacetime  \cite{Polchinski+}.

In the curved space of General Relativity, when the metric is dynamical variable, the global translational invariance is substituted by local spacetime diffeomorphism   invariance $SDiff_D$ and $\hat{x}^\mu (\xi)$ becomes, roughly speaking,  the Goldstone fields for $SDiff_D$. As for any Goldstone fields of the gauge symmetry, the degrees of freedom in coordinate functions can be gauged away (but not to zero) by imposing the `static gauge' conditions ({\it cf.} the `preferable gauge' $\phi=<\phi>$ above)
 \begin{eqnarray}\label{hx=gauge}
 \hat{x}^\mu (\xi) = ({\xi^m}
, 0,\ldots, 0)\;  \hspace{3cm} (\; \Leftrightarrow \qquad \phi=<\phi> \; )\; , \qquad \{\; {}^{\mu=0,1,\ldots, (D-1)\; ,} _{\quad m=0,1,\ldots, p\; ,} \quad
\end{eqnarray}
this is to say choose a local frame where $W^{(p+1)}$ looks locally as a flat hyperplane. In this gauge the  $p$--brane degrees of freedom are carried (are 'eaten') by the metric (or vielbein) gauge field.

The action describing the interaction of the gauge and Goldstone fields ({\it cf.} (\ref{LHiggs})) reads
\begin{eqnarray}\label{SEH+Sp}
 & S = {1\over 2\kappa}\int d^D x \, \sqrt{|g|} {\cal R}  +  S_{p}
 \qquad \left(\; \Longleftrightarrow \qquad{\cal L}= -{1\over 4} F_{\mu\nu}F^{\mu\nu} -{1\over 2} D_{\mu}\varphi^s
D^{\mu} \varphi^s - V(\varphi^2)\right) \; , \qquad
\end{eqnarray}
where $\int d^D x \, \sqrt{|g|} {\cal R}$ is the standard ($D$-dimensional) Einstein-Hilbert action and
$S_p$ is the $p$-brane action. In the simplest case of absence of the spacetime and worldvolume gauge fields (bulk and worldvolume fluxes) this reads
\begin{eqnarray}\label{Sp=}
S_{p}=
{T_p}\int d^{p+1} \xi
\sqrt{|det (
\partial_m\hat{x}^\mu \partial_n \hat{x}^\nu
g_{\mu\nu}(\hat{x}))|} \qquad & \left(\; \Longleftrightarrow \qquad -{1\over 2} D_{\mu}\varphi^s
D^{\mu} \varphi^s - V(\varphi^2)\right)  \;   \qquad
\end{eqnarray}
($\;S_{0}=
{m}\int d\tau
\sqrt{|det (
\dot{\hat{x}}{}^\mu \dot{\hat{x}}{}^\nu
g_{\mu\nu}(\hat{x}(\tau)))|}\; $ for the case of massive particle, $p=0$, $T_0=m$).
Varying the action with respect to the spacetime metric we arrive at the Einstein equation
\begin{eqnarray}\label{EiEq0}
 {\cal G}_{\mu\nu}(g):= \sqrt{|g|}
({\cal R}_{\mu\nu}- {1\over 2} g_{\mu\nu}{\cal R}) =
\kappa \,T_{\mu\nu} \;\;
\end{eqnarray}
with ${\cal R}=g^{\mu\nu}{\cal R}_{\mu\nu}$, ${\cal R}_{\mu\nu}= {\cal R}_{\mu\rho\, \nu}{}^{\rho}$ and the energy momentum tensor
\begin{eqnarray}\label{Tmn=}
 & T^{\mu \nu} = {T_p \over 4}
\int d^{p+1}\xi \sqrt{\gamma} \gamma^{mn}
\partial_m{\hat{x}}^\mu \partial_n{\hat{x}}^\nu
\delta^{D} (x-\hat{x}(\xi))
 \end{eqnarray}
having support on the $p$--brane worldvolume $W^{p+1}$ (\ref{WinM}).
Varying (\ref{SEH+Sp}) with respect to $\delta \hat{x}^\mu(\xi^m)$ one obtains the so-called  minimal surface equation ($\gamma_{mn}$:=$\partial_m\hat{x}^\mu
\partial_n\hat{x}^\nu g_{\mu\nu}(\hat{x})$ is the induced metric) \begin{eqnarray}\label{pgeod}
 \partial_{m} (\sqrt{|\gamma|}\gamma^{mn}g_{\mu\nu}(\hat{x})
\partial_{n}{\hat{x}}^\nu (\xi))   -{1\over 2} \sqrt{|\gamma|}\gamma^{mn} \partial_{m}{\hat{x}}^\nu
\partial_{n}{\hat{x}}^\rho (\partial_{\mu} g_{\nu\rho})(\hat{x})=0\; .  \qquad
 \end{eqnarray}
For p=0 case this can be written as the standard geodesic equation  ${d^2 \hat{x}^\mu\over ds^2} + \Gamma^{\mu}_{\nu \rho} \,
{d\hat{x}^\nu \over ds} \, {d\hat{x}^\rho\over ds}=0$.

In the complete analogy with the internal gauge symmetry case, Eq. (\ref{pgeod}) are dependent.
Indeed, as it is well known, the Bianchi identities $R_{[\mu\nu\, \rho]}{}^\sigma=0$ for the Riemann curvature tensor $R_{\mu\nu\, \rho}{}^\sigma$ result in the covariant conservation of the energy--momentum tensor,
$T^{\mu}{}_{\nu ; \mu}:= \partial_\mu (T^{\mu\rho}g_{\rho \nu})- {1\over 2}
T^{\mu\rho}\partial_\nu g_{\rho \mu} =0$. In our case this results in an equation with support on $W^{p+1}$; integrating it with a probe function one can show
(see \cite{Gursey} for the bosonic string and \cite{BdAI=PRD02} for super--$p$--brane cases) that this gives (\ref{pgeod}).

This implies that one can fix the gauge (\ref{hx=gauge}) and obtain the gauge fixed version of (\ref{pgeod}) from the gauge fixed version of Einstein equation, Eq. (\ref{EiEq=gauge}) which is  Eq. (\ref{EiEq0}) with
\begin{eqnarray}\label{Tp-st}
& T^{\mu\nu} = {T_p\over 4} \sqrt{|g^{[p+1]}|}\,g^{[p+1]mn}(\xi, \vec{0})
\delta_m^\mu\delta_n^\nu \delta^{(D-p-1)}(x^I)\, , \qquad \cases{m=0,1,\ldots, p\; , \cr I=p+1, \ldots , (D-1)\; .} \quad
\end{eqnarray}
Thus the $p$-brane source term given by energy momentum tensor (\ref{Tp-st}) in the {\it r.h.s.} of the Einstein equation (\ref{EiEq=gauge}) is the counterpart of the gauge field mass term in (\ref{DF=mA}). However \cite{BdAI=PRD02}--\cite{IB03-ProcsJdA}, this $p$-brane source term cannot be considered as a mass  term for graviton, but rather as a counterpart of the cosmological constant term which cannot be identified with the graviton mass \cite{IB03-ProcsJdA}.

\section{Spacetime filling brane ($p$--brane with $p=D-1$) as cosmological constant}
In the $p$=$D$-$1$ case  the $S_{p= D-1}=
 \propto {T_{D-1}}\int d^{D} \xi
\sqrt{|det (
\partial_m\hat{x}^\mu \partial_n \hat{x}^\nu
g_{\mu\nu}(\hat{x}))|}$ term in the interacting action (\ref{SEH+Sp}), can be reduced  (by gauge fixing $\xi^m = \delta_\mu^m x^\mu)$) to cosmological constant term
\begin{eqnarray}\label{Scosm=}
& S_{p=(D-1)}=
  {\Lambda\over \kappa}\int d^{D} x
\sqrt{|
g_{\mu\nu}({x}))|}\; ,
\end{eqnarray}

We have to stress that the notion of spacetime filling brane becomes nontrivial in the String/M-theory context (the references can be found in \cite{BdAI=PRD02}--\cite{IB03-ProcsJdA}) in particular, because such branes ({\it e.g.} D$9$--branes of type IIB string theory), are carriers of the gauge fields and fermionic Goldstone fields corresponding to spontaneous breaking of bulk supersymmetry.

With this in mind, {\it it is tempting to speculate on} that this simple observation can be useful as a basis to resolve the {\it cosmological constant problem}. Namely, the small value of cosmological constant with respect to a straightforward QFT estimation for vacuum energy, as well as the fact that its value is  nonzero and positive while supersymmetry and supergravity prefer zero or negative cosmological constant, might be explained just by that the cosmological constant is determined by the tension of a spacetime filling brane, $\Lambda \propto T_{D-1}$. However, developing such a conjecture goes beyond the score of this contribution.

A simple proof of that the cosmological constant cannot be considered as graviton mass has been presented in \cite{IB03-ProcsJdA}. Schematically, the arguments are as follows. In the case of small but non-vanishing cosmological constant,  $\Lambda \rightarrow 0$ but $\Lambda\not=0$, one may consider decomposition  over the flat spacetime $g_{\mu\nu}=\eta_{\mu\nu} + h_{\mu\nu}$ of the Einstein equation with cosmological constant,
\begin{eqnarray}\label{EiEq=h}
{\cal G}_{\mu\nu}(\eta + h) =  {\cal G}^{\eta}_{\mu\nu}(h) + {\cal O}(hh)= \Lambda \eta_{\mu\nu} + \Lambda h_{\mu\nu} \; . \qquad \end{eqnarray}
Here ${\cal G}^{\eta}_{\mu\nu}(h)$ contains the contributions of the first order in weak field $h$ and
the terms of higher order in $h$ are denoted by ${\cal O}(hh)$ (the Einstein tensor calculated with the flat metric  vanishes ${\cal G}_{\mu\nu}(\eta)=0$). The first impression might be that $\Lambda \eta_{\mu\nu}$ provides the mass term for graviton $h_{\mu\nu}$. This, however, is not the case. Indeed, if the first order approximation to Eq. (\ref{EiEq=h}) is given by ${\cal G}_{\mu\nu}(h)= \Lambda h_{\mu\nu}$ then zero order approximation should be $\Lambda \eta_{\mu\nu}=0$ which implies the vanishing of the cosmological constant $\Lambda =0$ (in contradiction with the original assumption). Thus the selfconsistent weak field approximation to (\ref{EiEq=h}) with $\Lambda\not=0$ can be formulated only assuming that the cosmological constant is of the same order as $h$, $h\propto \Lambda $, so that the first order approximation is given by ${\cal G}^\eta_{\mu\nu}(h)- \Lambda \eta_{\mu\nu}=0$; this equation does not contain mass term for $h$.

\section{Discussion and conclusion}

The deep reason beyod the fact that cosmological constant contribution to field equations cannot be considered as a mass term is that the equation possesses the gauge symmetry with
the same number of parameters as in the case of vanishing cosmological constant,
and, hence, the graviton maintains the same number of polarizations $D(D-3)/2$ (which
is $2$ for $D=4$) as in the case of massless spin 2 field. So the AdS graviton is also massless (like the Minkowski one).

Indeed, despite fixing the gauge (\ref{hx=gauge}), the functional (\ref{Scosm=}) and the corresponding $p=D$ interacting action (\ref{SEH+Sp}) is invariant under the spacetime diffeomorphism symmetry $Diff_D$. The reason for this is that the complete group of the gauge invariance of the interacting action (\ref{SEH+Sp}) is the direct product $Diff_D\otimes Diff_{p+1}$ of the spacetime and worldvolume diffeomorphism transformations. These correspond to the parametric functions $a^\mu(x)$ and $\delta \xi^m\, (\xi)$ in Eq.   (\ref{vdx=}) which  can be used to fix (on the worldvolume $W^{p+1}$) the gauge (\ref{hx=gauge}). This remains invariant under combined (spacetime and worldvolume) $Diff_{p+1}$ diffeomorphisms.

In the case of spacetime filling brane $p=D-1$ this residual $Diff_{p+1}$  invariance is the spacetime diffeomorphism invariance $Diff_{D}$ of the Einstein action with cosmological constant. Then the number of polarization of graviton remains the same as in the case of absence of cosmological constants. In the case of not-spacetime--filling brane, $p<D-1$, the gauge symmetry is reduced to $Diff_{p+1}$, but only on the worldvolume $W^{p+1}$,
while outside $W^{p+1}$ the $Diff_{D}$ still acts on graviton. Thus after gauge fixing graviton acquires some additional degrees of freedom (counterparts of additional polarizations), but only  on $W^{p+1}\in M^D$, the points of which for $p< (D-1)$ form the set of measure zero with respect to the set of all spacetime points. This is why one cannot identify the $p$--brane energy momentum tensor (\ref{Tp-st}) with a mass term for graviton.

The above discussion allows us to understand the difference of the gravity plus brane system with the model for massive spin 2 fields proposed not so long ago by G. 't Hooft \cite{t'Hooft=07},
\begin{eqnarray}\label{S'tHooft=} & S_{'t\;Hooft}= {1\over 2\kappa}\int d^D x \, \sqrt{|g|} {\cal R} + \int d^4x \sqrt{|g|}g^{\mu\nu}(x)\partial_\mu X^a(x)
\partial_\nu X^a(x) \; ,  \qquad a=1,\ldots, D \end{eqnarray}
Although seemingly similar to the spacetime filling brane action ((\ref{Sp=})  of (\ref{SEH+Sp})), the second term of  (\ref{S'tHooft=}) differs from that by that it involves not one but two spacetime metrics: $g_{\mu\nu}(x)$ and the constant $\delta_{ab}$ used to contract the indices of the Goldstone fields $X^a(x)$. As a result, the (spontaneous) breaking of the diffeomorphism symmetry by the dynamical system described by the action (\ref{S'tHooft=}) is complete,
the gauge fixed action ($ X^a= \delta_\mu^a x^\mu \cdot {m\over 2\kappa}$) \begin{eqnarray}\label{S'tHooft=gauge} & S_{'t\;Hooft}\vert_{_{gauge}}=  {1\over 2\kappa}\int d^D x \, \sqrt{|g|} {\cal R} +  {m^2\over 2\kappa}  \int d^4x \sqrt{|g(x)|}g^{\mu\mu}(x)\;   \qquad  \end{eqnarray}  does not possess gauge symmetry and, hence, the model describes a massive spin 2 field \cite{t'Hooft=07}.

As far as the $p$-brane plus gravity interacting system  described by Eqs. (\ref{SEH+Sp}) and  (\ref{Sp=}))  is concerned,
although the p-brane source terms play in it the same role as the mass terms of the gauge fields in the standard Higgs effect, they cannot be identified with the mass term, but rather with the counterpart of the cosmological constant term localized on the p-brane worldvolume \cite{BdAI=PRD02}--\cite{IB03-ProcsJdA}. The cosmological term itself can be associated with a spacetime feeling brane ($p=(D-1)$) and, as we have discussed above,  cannot be considered as a mass term.


\section*{References}
\begin{thebibliography}{9}

\bibitem{Higgs:1964ia}
  F.~Englert and R.~Brout,
``Broken symmetry and the mass of gauge vector mesons,''
  Phys.\ Rev.\ Lett.\  {\bf 13} (1964) 321;
 \\
P.~W.~Higgs,
  ``Broken symmetries, massless particles and gauge fields,''
  Phys.\ Lett.\  {\bf 12} (1964) 132.



\bibitem{Duff=PRD75}
M.J. Duff,
``Dynamical breaking of general covariance and massive spin-2 mesons,''
Phys. Rev. {\bf D12}, 3969-3971 (1975);
%
C.~Omero and R.~Percacci,
  ``Generalized Nonlinear Sigma Models In Curved Space And Spontaneous
  Compactification,''
  Nucl.\ Phys.\   {\bf B165} 351 (1980);
  R.~Percacci,
  ``The Higgs Phenomenon in Quantum Gravity,''
  Nucl.\ Phys.\  B {\bf 353}, 271 (1991)
  [arXiv:0712.3545 [hep-th]];
  M.~Porrati,
  ``Higgs phenomenon for 4-D gravity in anti de Sitter space,''
  JHEP {\bf 0204}, 058 (2002)
  [arXiv:hep-th/0112166].
  A.~H.~Chamseddine,
  ``Spontaneous symmetry breaking for massive spin-2 interacting with
  gravity,''
  Phys.\ Lett.\  B {\bf 557}, 247 (2003)
  [arXiv:hep-th/0301014].

\bibitem{t'Hooft=07} G. 't Hooft, ``Unitarity in the Brout-Englert-Higgs Mechanism for Gravity,'' arXiv:0708.3184 [hep-th].

\bibitem{BdAI=PRD02}
  I.~A.~Bandos, J.~A.~De Azcarraga and J.~M.~Izquierdo,
  ``Supergravity interacting with bosonic p-branes and local supersymmetry,''
  Phys.\ Rev.\  D {\bf 65}, 105010 (2002)
  [arXiv:hep-th/0112207].


\bibitem{BdAIL=PRD03}
  I.~A.~Bandos, J.~A.~de Azcarraga, J.~M.~Izquierdo and J.~Lukierski,
  ``D = 4 supergravity dynamically coupled to a massless superparticle in a
  superfield Lagrangian approach,''
  Phys.\ Rev.\  {\bf D67}, 065003 (2003)
  [arXiv:hep-th/0207139];
  ``On dynamical supergravity interacting with super-p-brane sources,''
 in {\it 3rd International Sakharov  Conference, Moscow, June 24-29,
 2002. Proceedings}, Scientific World Publ.
 Co, Edts. A. Semikhatov et. al., 2003 Vol. 2, pp. 312-324
[arXiv:hep-th/0211065]; \\
  I.~A.~Bandos and J.~M.~Isidro,
 ``D = 4 supergravity dynamically coupled to superstring in a superfield
  Lagrangian approach,''
  Phys.\ Rev.\   {\bf D69}, 085009 (2004)
  [arXiv:hep-th/0308102].





\bibitem{BdAIL=Higgs03}
  I.~A.~Bandos, J.~A.~de Azcarraga, J.~M.~Izquierdo and J.~Lukierski,
  ``Gravity, p-branes and a spacetime counterpart of the Higgs effect,''
  Phys.\ Rev.\  {\bf D68}, 046004 (2003)
  [arXiv:hep-th/0301255].





\bibitem{IB03-ProcsJdA}
  I.~A.~Bandos,
  ``Supergravity interacting with superbranes and spacetime Higgs effect in
  general relativity,'',
 in: {\it Symmetries in gravity and field theory.} {\sl Conference
  dedicated to 60th birthday of Jos\'e Adolfo de Azc\'arraga.
Salamanca. June 9-11, 2003.} (eds. J.M. Cercer\'o, Y. Pilar
 Grac\'{\i}a), Salamanca 2003, pp. 489--504
[arXiv:hep-th/0312095].




\bibitem{IB+JdA=JHEP05}  I.~A.~Bandos and J.~A.~de Azcarraga,
  ``Dirac equation for the supermembrane in a background with fluxes from a
  component description of the D = 11 supergravity-supermembrane  interacting
  system,''
  JHEP {\bf 0509}, 064 (2005)
  [arXiv:hep-th/0507197].


\bibitem{Polchinski+}
J.~Hughes and J.~Polchinski,
  ``Partially broken global supersymmetry and the superstring,''
  Nucl.\ Phys.\  {\bf B278}, 147--169 (1986); \\
J.~Hughes, J.~Liu and J.~Polchinski,
  ``Supermembranes,''
  Phys.\ Lett.\   {\bf B180}, 370--374 (1986).

\bibitem{Gursey}
M. G\"urses and F. G\"ursey,
 ``Derivation of the string equations of motion in general relativity,''
Phys.\ Rev.\ {\bf D11}, 967-969
(1975);
\\
C. Aragone and S. Deser,
 ``String dynamics from energy--momentum conservation,''
{Nucl. Phys.} {\bf B92}, 327-333
(1975).



\end{thebibliography}
\end{document}